\documentclass{jors}

\usepackage{lineno,hyperref}

\usepackage{amsmath}
\usepackage{amssymb}
\usepackage{mathtools}
\usepackage{array}
\usepackage{tabu}
\usepackage{graphicx}
\usepackage{subcaption}
\usepackage{geometry}
\usepackage{lipsum}
\usepackage{longtable}
\usepackage{pdflscape}
\usepackage{adjustbox}
\usepackage{float}
\usepackage[utf8]{inputenc}
\usepackage{lscape}
\usepackage{multirow}
\usepackage{gensymb}
\usepackage{color, colortbl}
\usepackage{soul}

\definecolor{Gray}{gray}{0.95}
\newcolumntype{g}{>{\columncolor{Gray}}c}
\usepackage{sectsty}
\sectionfont{\large}
\subsectionfont{\normalsize}
\subsubsectionfont{\normalsize}
\paragraphfont{\normalsize}

\usepackage{titlesec}
\titlespacing\section{0pt}{12pt}{6pt}
\titlespacing\subsection{0pt}{12pt}{6pt}
\titlespacing\subsubsection{0pt}{12pt}{6pt}

\usepackage[numbers]{natbib}
\usepackage{notoccite}
\begin{document}

{\Large \bf PowNet: a power systems analysis model for \\large-scale water-energy nexus studies} \\

{AFM Kamal Chowdhury\textsuperscript{1}, Jordan Kern\textsuperscript{2}, Thanh Duc Dang\textsuperscript{1}, Stefano Galelli\textsuperscript{1}}\\ \\
{\textsuperscript{1} Pillar of Engineering Systems and Design, Singapore University of Technology and Design, Singapore 487372}\\
{\textsuperscript{2} Department of Forestry and Environmental Resources, North Carolina State University, Jordan Hall Addition, 2820 Faucette Dr., NC 27695}\\


Corresponding author: AFM Kamal Chowdhury (chowdhury@sutd.edu.sg)

\rule{\textwidth}{1pt}

\section*{Abstract}
PowNet is a free modelling tool for simulating the Unit Commitment / Economic Dispatch of large-scale power systems. PowNet is specifically conceived for applications in the water-energy nexus domain, which investigate the impact of water availability on electricity supply. To this purpose, PowNet is equipped with features that guarantee accuracy, reusability, and computational efficiency over large spatial and temporal domains. Specifically, the model (i) accounts for the techno-economic constraints of both generating units and transmission networks, (ii) can be easily coupled with models that estimate the status of generating units as a function of the climatic conditions, and (iii) explicitly includes import/export nodes, which are often found in cross-border systems. PowNet is implemented in Python and runs with the help of any standard optimization solver (e.g., Gurobi, CPLEX). Its functionality is demonstrated on the Cambodian power system.

\section*{Keywords} Unit commitment; economic dispatch; transmission networks; water-energy nexus; energy systems; power systems; Python

\rule{\textwidth}{1pt}

\section*{(1) Overview}

\vspace{0.5cm}

\section{Introduction} \label{intro}


Fuelled by economic and population growth, electricity demand is rapidly increasing in many parts of the world. At the same time, several countries are cutting carbon dioxide emissions through a larger dependance on renewable resources (e.g., hydro, wind and solar) \cite{GIELEN201938}. Yet, electricity supply from these resources varies over multiple temporal scales---from hourly to seasonal and inter-annual---thereby requiring more flexible power grids \cite{WELSCH2014600,SU2017172}. These challenges have prompted the development of several modelling tools for energy and electricity systems; see \cite{RINGKJOB2018440} for a comprehensive review. Broadly speaking, models are used for two main tasks, long-term planning (e.g., capacity expansion, investment optimization) and short-term management (e.g., power flow analysis, unit commitment). Hereafter, we refer to these two groups as \textit{energy} and \textit{power systems} analysis models \cite{brown2018pypsa}. \\

Importantly, modelling tools are used not only for `traditional' tasks in the energy and power domains, such as investment optimization and grid analysis, but also for interdisciplinary research spanning across multiple engineering domains. For example, models are increasingly being used in studies on the water-energy nexus, which aim to characterize the interdependencies between these two critical sectors and introduce planning and management solutions that span across coupled systems \cite{DAI2018393}. The strategic importance of energy and power models is thus likely to increase in the near future, as climatic changes are expected to transform both supply and demand of electricity \cite{vanVliet_2016,TURNER2017663,Turner2019}. \\ 

An important application in the water-energy nexus domain is Unit Commitment / Economic Dispatch (UC/ED), in which a computer model is used to optimize the operations of power generating units---more precisely, UC determines when and which generating units to start-up and shut-down, while ED establishes the amount of power supplied by each unit \cite{Conejo2018}. The correct simulation of the UC/ED decision-making process is indeed a critical step if one seeks to decipher the effect of water availability on the performance of a power system. To this purpose, researchers and practitioners typically rely on power systems models---such as PROMOD \cite{Voisin2016} or GENESYS \cite{Turner2019}---that account for the techno-economic constraints of the generating units, but often adopt simplified representations of the transmission networks; an assumption that may lead to a misrepresentation of power system's performance \cite{DEANE2012303,SAVVIDIS2019503}. To explicitly account for electricity transmission in a UC/ED exercise, one can only rely on a handful of models, namely PSAT \cite{PSAT}, PyPSA \cite{brown2018pypsa}, and PLEXOS \cite{OCONNELL2019745}. Yet, the latter is a proprietary software, and thereby not freely available to researchers. \\

Here, we contribute to this growing field and introduce PowNet, a UC/ED modelling tool specifically conceived for applications in the water-energy nexus domain. PowNet builds on the UC/ED model first presented by \cite{kern2017}, and complements it with a comprehensive representation of the high-voltage transmission lines. As we shall see later, PowNet has a few desirable features. First, it bases the power flow calculation on a Direct Current (DC) network; a choice that strikes a reasonable balance between modelling accuracy and data and computational requirements. This is particularly important for water-energy applications, which are generally carried out over extensive spatial and temporal domains. Moreover, it should be noted that the error introduced by DC flow is generally negligible (as compared to a full-scale power system model), except for high loadings with increased reactive power \cite{Stott2009,Brown2016}. To minimize such instances, PowNet incorporates the N-1 criterion, which leaves part of the lines' capacity unused, thereby allowing for reactive power flows \cite{Brown2016,Schlecht20141}. Second, PowNet accounts for the potential effect of water availability on the UC/ED process. Specifically, the user can soft link PowNet with any model able to process climatic data and provide information on the status of renewable and non-renewable resources. Third, PowNet explicitly includes the export/import nodes---as substations and generators--which are key components of cross-borders interconnections. Fourth, PowNet is fully written in Python, a programming language that should enhance its dissemination and reusability. \\

The next section describes the mathematical model on which PowNet is based, while its Python implementation and architecture are described in Section 3. Section 4 presents the quality control, along with a sample implementation for the Cambodian power system. The concluding remarks are outlined in Section 5.

\section{Functionality}\label{func}

In PowNet, a power system is represented by a set of nodes ($n\,\epsilon\,N$) that include power plants and high-voltage substations. Given a pre-defined planning horizon (e.g., 24 hours), the model (1) schedules the operating status of the dispatchable power plant units (e.g., coal, gas, oil, and biomass), and (2) determines the hourly dispatch of electricity from the committed units and variable renewable resources (e.g., hydro, solar, wind) that meets the system's demand at a minimum cost. As illustrated in Figure 1, the electricity available through the variable renewable resources is an input that can be obtained from existing data or simulated by a separate model (e.g., a hydrological-hydraulic model for the available hydropower). PowNet can also include import stations as dispatchable units. The scheduling and dispatch of hourly electricity is constrained by the design features of the power plants as well as the capacity and susceptance of the transmission lines. Overall, PowNet solves a Mixed Integer Linear Program (MILP), whose objective function, decision variables, and constraints are described next (the notation used in the following sections is summarized in Table 1). 

\begin{figure}[h]
	\centering
	\includegraphics[trim={1.0cm 3.0cm 1.0cm 3.4cm},clip=true,width=1.0\linewidth]{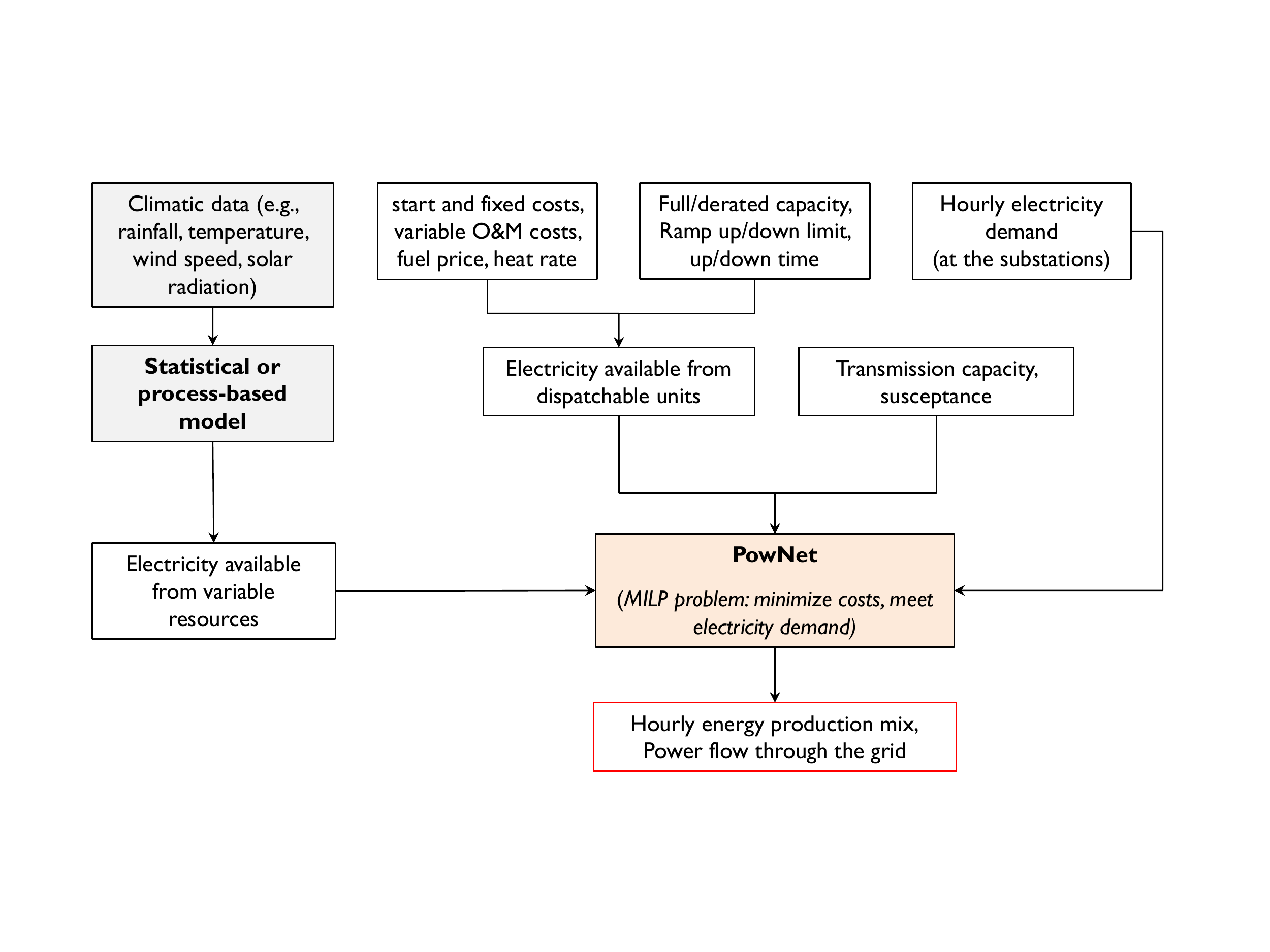}  
	\caption[watenmodel]{Graphical representation of PowNet. Black and red boxes represent input and output elements, respectively. Grey boxes represent elements not included in the model.}
	\label{fig:pownet}
\end{figure}

\subsection{Objective function and decision variables}

The goal of PowNet is to meet the hourly electricity demand at the substations while minimizing the production cost over a planning horizon, say 24 hours ($t\,\epsilon\,T$). The production cost of the dispatchable units ($g\,\epsilon\,G$) depends on their fixed operation and maintenance (O\&M) costs (${FixedCost}_{g}$), start-up cost (${StartCost}_{g}$), variable costs for heat rate (${HeatRate}_{g}$) and fuel price (${FuelPrice}_{g}$), and variable O\&M costs (${VarCost}_{g}$). The shut-down costs of the units are generally negligible when compared to the other costs \cite{Truby2014}, and are thus not considered in our model. The cost of imported electricity depends on the import price (${ImportPrice}_{i}$), which is specified for each import source ($i\,\epsilon\,I$). The production cost of the $rn$-th variable renewable resource ($rn\,\epsilon\,RN$) depends on the unit production cost (${UnitCost}_{rn}$), which is typically smaller than the one of the dispatchable unit \cite{kern2017}. The objective function is thus formulated as follows:
\begin{linenomath}
	\begin{align}\label{eqobj}
	\begin{split}
	& \min \sum_{t=1}^{T} \Big( \sum_{g}^{G}\big({FixedCost}_{g} \times {ON}_{g,t} + {Elec}_{g,t} \times ({HeatRate}_{g} \times \\
	& {FuelPrice}_{g} + {VarCost}_{g}) + {StartCost}_{g} \times {Switch}_{g,t}\big) +\\
	& \sum_{i}^{I}({Elec}_{i,t} \times {ImportPrice}_{i}) +\sum_{rn}^{RN}({Elec}_{rn,t} \times {UnitCost}_{rn}) \Big),	
	\end{split}
	\end{align}	
\end{linenomath}
where ${ON}_{g,t}$ and ${Switch}_{g,t}$ are two decision variables describing the operational status (on-line or off-line, and to be started-up or shut-down) of the $g$-th dispatchable unit at hour $t$, while ${Elec}_{g,t}$, ${Elec}_{i,t}$, and ${Elec}_{rn,t}$ denote the hourly electricity production of the $g$-th dispatchable unit, $i$-th import source, and $rn$-th  variable renewable resource, respectively. At each hour \textit{t}, PowNet optimizes a few additional decision variables, namely, the voltage angle ${VoltAngle}_{n,t}$ at each node $n$ of the network (needed to estimate the transmission through each line), and the spinning and non-spinning reserves ${SpinRes}_{g,t}$ and ${NonSpinRes}_{g,t}$.

\begin{table}[]
	\small\addtolength{\tabcolsep}{18pt}
	\centering
	\caption{Notations used to describe PowNet.}
	\label{TabNotation}
	\begin{adjustbox}{width=1\textwidth}
		\begin{tabular}{ll}
			\hline
			\multicolumn{2}{l}{\textbf{Indices and sets:}}                                  \\
			\textit{n}   & any node in the system                                       \\
			\textit{N}   & set of all nodes in the system                               \\
			\textit{g}   & dispatchable unit                                  \\
			\textit{G}   & set of all dispatchable units                                \\
			\textit{i}   & import node                                                  \\
			\textit{I}   & set of all import nodes                                      \\
			\textit{rn}  & variable renewable resource (e.g., hydro, wind, solar)       \\
			\textit{RN}  & set of all variable renewable resources                            \\
			\textit{k}   & sink node connected by the transmission line to any node \textit{n}  \\
			\textit{t}   & time-step (hour, h)                                          \\
			\textit{T}   & planning horizon (e.g., 24 h)                                 \\
			\textit{}    &                                                              \\
			\multicolumn{2}{l}{\textbf{Parameters of the dispatchable units (\textit{g}):}} \\
			\textit{${MaxCap}_{g}$}    & maximum capacity (MW)                       \\
			\textit{${MinCap}_{g}$}    & minimum capacity (MW)                         \\
			\textit{${FixedCost}_{g}$}    & fixed O\&M cost (\$)                 \\
			\textit{${StartCost}_{g}$}    & start-up cost (\$/start)                    \\
			\textit{${HeatRate}_{g}$}    & heat rate (MMBtu/MWh)                       \\
			\textit{${FuelPrice}_{g}$}    & price of fuel (\$/MMBtu)                   \\
			\textit{${VarCost}_{g}$}    & variable O\&M cost (\$/MWh)                 \\
			\textit{${Ramp}_{g}$}    & ramping limit (MW/h)                             \\
			\textit{${MinUpTime}_{g}$}    & minimum up time (h)                        \\
			\textit{${MinDownTime}_{g}$}    & minimum down time (h)                     \\
			\textit{}    &                                                              \\
			\multicolumn{2}{l}{\textbf{Parameters of the import nodes (\textit{i}):}}    \\
			\textit{${MaxCap}_{i}$}    & maximum allowable import (MW)                    \\
			\textit{${ImportPrice}_{i}$}    & price of imported electricity (\$/MWh)      \\
			\textit{}    &                                                              \\
			\multicolumn{2}{l}{\textbf{Parameters of the transmission network:}}                     \\
			\textit{${LineSus}_{n,k}$}   & susceptance (Siemens) of the transmission line between nodes \textit{n} and \textit{k} \\
			\textit{${LineCap}_{n,k}$}   & capacity (MW) of the transmission line between nodes \textit{n} and \textit{k} \\
			\\
			\multicolumn{2}{l}{\textbf{Input time series (hourly):}}                     \\
			\textit{${RenewAvail}_{rn,t}$}   & available electricity (MWh) from the \textit{rn}-th renewable resource   \\
			\textit{${Demand}_{n,t}$}   & electricity demand (MWh) (or export) at any node \textit{n}   \\
			\\
			\multicolumn{2}{l}{\textbf{Decision variables (at each hour \textit{t}):}}                     \\
			\textit{${ON}_{g,t}$}   & binary (0,1) variable indicating if unit \textit{g} is online (1) or offline (0) \\
			\textit{${Switch}_{g,t}$}   & \multirow{2}{*}{\begin{tabular}[c]{@{}l@{}} binary (0,1) variable indicating if unit \textit{g} must be started-up\\ (${Switch}_{g,t}$=1 only when ${ON}_{g,t-1}$=0 is followed by ${ON}_{g,t}$=1) \end{tabular}}\\
			&\\
			\textit{${Elec}_{g*,t}$}   & electricity (MWh) generated by dispatchable unit \textit{g} (*or any other powerplant)   \\
			\textit{${VoltAngle}_{n,t}$}   & voltage angle (radian) at any node \textit{n} \\
			\textit{${SpinRes}_{g,t}$}   & spinning reserve (MWh) at unit \textit{g} \\
			\textit{${NonSpinRes}_{g,t}$}   & non-spinning reserve (MWh) at unit \textit{g} \\
			\hline			
		\end{tabular}
	\end{adjustbox}
\end{table}

\subsection{Constraints} \label{constraints}
The scheduling and dispatch of hourly electricity is subject to multiple constraints accounting for the technical specifications of generating units (i.e., capacity, minimum up/down time, ramping limit), energy balance at each node (demand and supply), capacity and susceptance of the transmission lines, transmission loss and N-1 criterion, and minimum requirements of spinning and non-spinning reserves.

\paragraph{\normalsize Logical constraints.}

Similarly to \cite{kern2017}, we adopt a set of logical constraints to bind the operational status of the various power plant units. First of all, we introduce a constraint (equation (\ref{eqlogic})) on the operational status of the dispatchable units, according to which an off-line unit can be started up but not shut down, and vice versa:
\begin{linenomath}
	\begin{align}\label{eqlogic}
	\begin{split}
	& {Switch}_{g,t} \geq 1 - {ON}_{g,t-1} - (1 - {ON}_{g,t}); \\
	& {ON}_{g,t}\,\epsilon\,\{0,1\},\,{Switch}_{g,t}\,\epsilon\,\{0,1\},\,\forall_{g},\,\forall_{t},
	\end{split}
	\end{align}
\end{linenomath}
where, the binary variable ${ON}_{g,t}$ determines whether the $g$-th dispatchable unit is off-line (or on-line) at hour $t$, while ${Switch}_{g,t}$ indicates whether an off-line unit must be started up (or not). \\	

The operational status of each unit is also constrained by the minimum up and down time (${MinUpTime}_{g}$ and ${MinDownTime}_{g}$) required to start it up or shut it down, that is:
\begin{linenomath}
	\begin{align}\label{eqminup}
	\begin{split}
	& {ON}_{g,t} - {ON}_{g,t-1} \leq {ON}_{g,j}; \\
	& \forall_{g},\, t\,\epsilon\,\{2\:to\:(T-1)\},\,t<j\leq \min(t+{MinUpTime}_{g}-1,T),
	\end{split}
	\end{align}	
	\begin{align}\label{eqmindn}
	\begin{split}
	& {ON}_{g,t-1} - {ON}_{g,t} \leq 1- {ON}_{g,j}; \\
	& \forall_{g},\, t\,\epsilon\,\{2\:to\:(T-1)\},\,t<j\leq \min(t+{MinDownTime}_{g}-1,T).
	\end{split}
	\end{align}	
\end{linenomath}
Over a planning horizon $T$, equations (\ref{eqminup}) and (\ref{eqmindn}) thus force PowNet to account for the minimum number of hours necessary to start the $g$-th unit up (or to shut it down) at hour $j$.

\paragraph{\normalsize Ramping limits.}

The ramping up and down hourly limits of the $g$-th unit (${Ramp}_{g}$) are constrained by equations (\ref{eqramup}) and (\ref{eqramdn}):
\begin{linenomath}
	\begin{align}\label{eqramup}
	\begin{split}
	& {Elec}_{g,t} - {Elec}_{g,t-1} \leq {Ramp}_{g}; \;\;\forall_{g},\,\forall_{t},
	\end{split}
	\end{align}	
	\begin{align}\label{eqramdn}
	\begin{split}
	& {Elec}_{g,t-1} - {Elec}_{g,t} \leq {Ramp}_{g}; \;\;\forall_{g},\,\forall_{t}.
	\end{split}
	\end{align}	
\end{linenomath}
In other words, equations (\ref{eqramup}) and (\ref{eqramdn}) ensure that the increase or decrease of the power generation during two consecutive hours (${Elec}_{g,t}$ and ${Elec}_{g,t-1}$) is below the ramping limit (${Ramp}_{g}$) of the $g$-th unit.

\paragraph{\normalsize Capacity constraints.}

Equation (\ref{eqgenlim}) accounts for the minimum and maximum capacity of the generating units. In particular, equation (\ref{eqgenlim}) indicates that the hourly electricity production ${Elec}_{g,t}$ from the $g$-th dispatchable unit is bounded by its minimum and maximum capacity (${MinCap}_{g}$ and ${MaxCap}_{g}$):  
\begin{linenomath}
	\begin{align}\label{eqgenlim}
	\begin{split}
	& {MinCap}_{g} \times {ON}_{g,t} \leq {Elec}_{g,t} \leq {MaxCap}_{g} \times {ON}_{g,t} \times {DerateF}_{g,t}; \;\;\forall_{g},\,\forall_{t}.
	\end{split}
	\end{align}	
\end{linenomath}

The term ${DerateF}_{g,t}$ is used to account for the impact of droughts on freshwater-dependent dispatchable units---during a prolonged dry spell, for example, a thermo-power unit may not be able to run at full capacity because of its limited cooling capability. The default value of ${DerateF}_{g,t}$ is one, but it can be modified to a smaller value if water availability becomes a limiting factor. Note that the value of ${DerateF}_{g,t}$ must be supplied by the user, which can rely on a variety of methods for such task \cite{Voisin2016,lubega2018maintaining}. \\

Similarly, the hourly electricity import ${Elec}_{i,t}$ from the $i$-th import source is constrained by the maximum allowable import (${MaxCap}_{i}$), as indicated by equation (\ref{eqimplim}):
\begin{linenomath}
	\begin{align}\label{eqimplim}
	\begin{split}
	& 0\leq {Elec}_{i,t} \leq {MaxCap}_{i} \times {ON}_{i,t}; \;\;\forall_{i},\,\forall_{t}. 
	\end{split}
	\end{align}	
\end{linenomath} \\

Finally, the amount of electricity ${Elec}_{rn,t}$ dispatched from the $rn$-th renewable resource (e.g., hydro, wind, and solar) is bounded by its availability (${RenewAvail}_{rn,t}$):
\begin{linenomath}
	\begin{align}\label{eqhydlim}
	\begin{split}
	& 0\leq {Elec}_{rn,t} \leq {RenewAvail}_{rn,t}; \;\;\forall_{rn},\,\forall_{t}.
	\end{split}
	\end{align}	
\end{linenomath}

The value of ${RenewAvail}_{rn,t}$ is generally modelled through the use of climatic data. For example, the available hydro-electricity can be estimated by feeding a hydrological-hydraulic model with rainfall and temperature data \cite{Dang2019}. Similarly, the electricity available in wind and solar power plants can be simulated by harnessing data on wind speed \cite{Papavasiliou} and solar radiation \cite{SAM2014}.

\paragraph{\normalsize Energy balance.}	

Equation (\ref{eqbal}) applies the energy balance at each node \textit{n} connected to any other node \textit{k} of the power system \cite{Conejo2018}. The left-hand side of equation (\ref{eqbal}) accounts for (1) the electricity inputs (${Elec}_{n,t}$) from the available power plants and/or import-sources connected to \textit{n}, and (2) the electricity used to meet domestic demand or export (${Demand}_{n,t}$ or ${Export}_{n,t}$) at the node. The right-hand side indicates that the electricity transferred (in or out) between \textit{n} and \textit{k} is proportional to the difference of the voltage angles at these nodes (${VoltAngle}_{n,t}$ and ${VoltAngle}_{k,t}$), where the susceptance of the transmission line (${LineSus}_{n,k}$) is the proportionality constant \cite{Bergen1999}. The voltage angle at an arbitrary reference node (usually, the node with highest demand) is set to zero, while the voltage angle at any other node can be positive or negative \cite{Conejo2018}. The parameter \textit{TransLoss} is used to discount the energy production, or import, by a given percentage, so as to account for the transmission losses (similar to \cite{GUERRA20161}). Equation (\ref{eqbal}) is formulated as:
\begin{linenomath}	
	\begin{align}\label{eqbal}
	\begin{split}
	& (1- TransLoss) \times \sum{Elec}_{n,t} - {Demand}_{n,t} \,(-{Export}_{n,t}) \\
	&= \sum_{k\,\epsilon\,N}{LineSus}_{n,k} \times \big({VoltAngle}_{n,t} -{VoltAngle}_{k,t}\big); \;\;\forall_{n},\,\forall_{t}.
	\end{split}
	\end{align}
\end{linenomath}	

The hourly demand (and export) time series at the substations represent an input to the model. These data can be obtained from observed records (see Section \ref{setup}) or modelled through relevant factors, such as population or temperature \citep{SU2017172}.

\paragraph{\normalsize Transmission capacity constraints.}

With equation (\ref{eqtranslim}), known as the N-1 criterion, PowNet limits the electricity transfer between any node-pair (\textit{n} and \textit{k}) below a certain percentage ($N1Criterion$) of the line-capacity (${LineCap}_{n,k}$), thereby leaving the remaining fraction as a safety margin:
\begin{linenomath}	
	\begin{align}\label{eqtranslim}
	\begin{split}
	& -N1Criterion \times {LineCap}_{n,k} \leq {LineSus}_{n,k} \times \big({VoltAngle}_{n,t} -{VoltAngle}_{k,t}\big)\\
	&\leq N1Criterion \times {LineCap}_{n,k}; \;\;\forall_{n},\,\forall_{t},\,k\,\epsilon\,N.
	\end{split}
	\end{align}
\end{linenomath}

A common value for the parameter $N1Criterion$ is 75\% (c.f., \cite{Schlecht20141}). As illustrated in Appendix A, the parameters ${LineCap}_{n,k}$ and ${LineSus}_{n,k}$ are estimated from design specifications of the transmission lines (e.g., size and length, voltage level, number of circuits and conductors, and capacity per circuit), which can be easily obtained from technical reports. 

\paragraph{\normalsize Electricity reserve.}

The last two equations ensure that (1) the hourly electricity reserve is larger than a predefined percentage of the system's demand at time $t$ (e.g., $ResMargin$ = 15\%, \cite{GUERRA20161}), and (2) the minimum spinning reserve is a predefined percentage of the total reserve (e.g., $SpinMargin$ = 50\%, \cite{kern2017}):
\begin{align}\label{eqres1}
\begin{split}
& \sum_{g}^{G}\big({SpinRes}_{g,t} + {NonSpinRes}_{g,t}\big) \geq ResMargin \times \sum_{n}^{N} {Demand}_{t}; \;\;\forall_{t},
\end{split}
\end{align}	
\begin{align}\label{eqres2}
\begin{split}
& \sum_{g}^{G}{SpinRes}_{g,t}  \geq SpinMargin \times ResMargin \times \sum_{n}^{N} {Demand}_{t}; \;\;\forall_{t}.
\end{split}
\end{align}	

Two additional constraints are used to ensure that spinning and non-spinning reserves are served by on-line and off-line generators, respectively. Furthermore, users can allocate all or some specific generators (e.g., oil-fired units) for the minimum reserve requirements.

\section{Implementation and architecture}


PowNet is implemented in three Python scripts, namely `PowNetModel.py', `PowNetDataSetup.py', and `PowNetSolver.ipynb'. `PowNetModel.py' contains the main model structure, with the objective function and constraints described in Section \ref{func}. The model structure is based on the Pyomo optimization package \cite{Hart2011}. `PowNetDataSetup.py' prepares the data required to execute the model. Specifically, it outlines the input data into several .csv files that are read as DataFrame objects. Then, the script generates a .dat file in which all data are specified in a format that is executable by Pyomo. Finally, `PowNetSolver.ipynb' executes the model and solves the optimization problem using a standard solver, such as Gurobi or CPLEX. The script also generates .csv files containing the value of each decision variable. \\

PowNet input data are categorized as: (i) parameters of the dispatchable units, (ii) parameters of the transmission lines, (iii) hourly time series of electricity demand at the substations, and (iv) hourly time series of electricity available from variable renewable resources. The techno-economic parameters of the dispatchable units are provided in a .csv file that also includes identification data, such as the name and type of each unit and the node to which they belong. The type of unit depends on fuel and turbine (e.g., `coal\_st' stands for a coal-fired unit with steam turbine). The techno-economic parameters are considered constant over the simulation period. In addition, the .csv file contains the value of the derating factor (see Section \ref{constraints}). The data concerning the parameters of the transmission lines (i.e., susceptance and capacity), hourly electricity demand, and electricity available from renewable resources are provided in separate .csv files. The data regarding the length of the simulation period, planning horizon, transmission loss, N-1 criterion, and hourly reserve margins are provided directly by the user in the script `PowNetDataSetup.py'.

\section{Quality control}


So far, PowNet has been tested on the Laotian \cite{PowNetLaos1,PowNetLaos2} and Thai \cite{PowNetThai} power systems. In both cases, the model output was validated against observed statistics, such as the seasonal or annual generation mix. Here, we present an implementation of PowNet for the Cambodian power system. The data and code used for this demonstration are available in GitHub along with step-by-step instructions on how to customize them. The next section presents an overview of the model setup, followed by an evaluation of its performance.

\subsection{Setup} \label{setup}

\begin{figure}[h]
	\centering
	\includegraphics[width=0.75\linewidth]{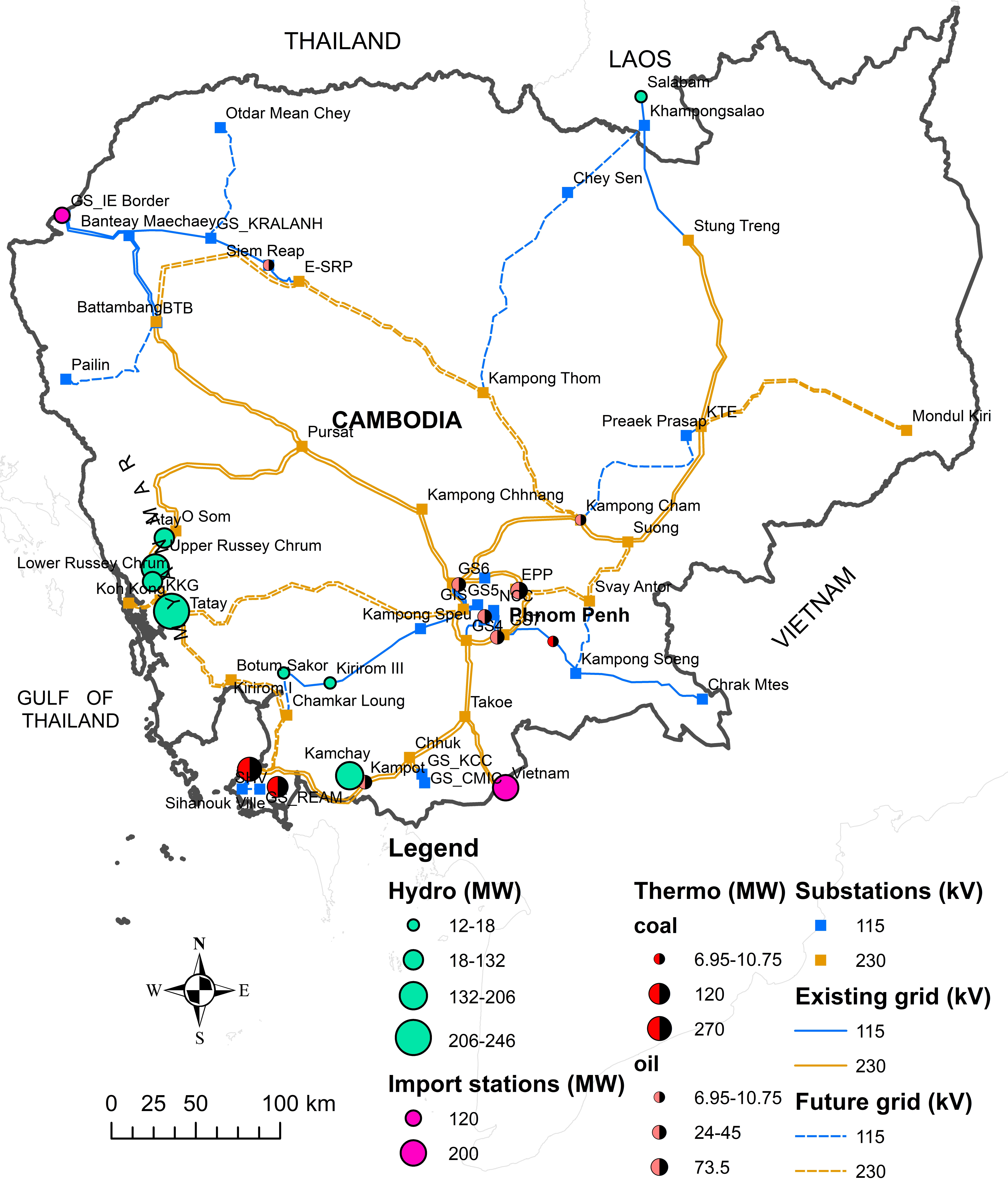}
	\caption{Main generation and transmission components of the Cambodian power system, as of 2016. The dashed lines denote components that are either planned or under construction.} 	
	\label{fig:cambgrid}
\end{figure}

Our implementation is based on the infrastructure built and operated in 2016. The system consists of 30 nodes---including power plants and substations---connected by high-voltage transmission lines (Figure \ref{fig:cambgrid}) \citep{EDC2016}. The total generating capacity is 2,000 MW, with the national peak demand equal to 1,192 MW. The domestic capacity includes three coal-fired units (400 MW), 15 oil-fired units (282 MW), and six hydropower plants (930 MW). The system is also connected to two import stations (dispatchable) from Vietnam (200 MW) and Thailand (120 MW), and a hydropower plant in Laos. To set PowNet, data on the parameters of dispatchable units and transmission lines were extracted from different technical reports \citep{Truby2014,EDC2016,EIA2016}. (Further details about the estimation of the transmission parameters are given in Appendix A.) We also extracted data on province-wise, monthly-varied peak electricity demand from \citep{EDC2016}. Spatially, the province-wise demand data were disaggregated to substations based on their voltage-levels. For the disaggregation of monthly data to hourly resolution, we used a week-daily and a clock-hourly demand profile to incorporate the variability of demand over days in a week and hours in a day, respectively. The time series of available hydropower was simulated by a conceptual hydrologic model (see Appendix B).\\

The model was run with a 24 hour planning horizon, threshold for the N-1 criterion equal to 75\% (of the line's capacity), and reserve margin equivalent to 15\% of the system's demand. The model was calibrated against the 2016 (observed) annual generation mix by tuning the fuel price value in Equation (1). For each day (24 hours) of simulation, PowNet optimizes around 2,600 variables (continuous and binary). The model was tested with two solvers (Gurobi and CPLEX) on an Intel(R) Core(TM) i7-8700 CPU @ 3.20 GHz with 8 GB RAM running Microsoft Windows 10. They both yield the same output, with slightly different computational requirements---$\sim$3.5 and $\sim$3.0 seconds per day of simulation, respectively. Further tests on a Linux operating system show similar performance.\\ 

\subsection{Performance}

\begin{figure}[h]
	\centering
	\includegraphics[trim={0 0.1cm 0 0.20cm},clip=true,width=0.8\linewidth]{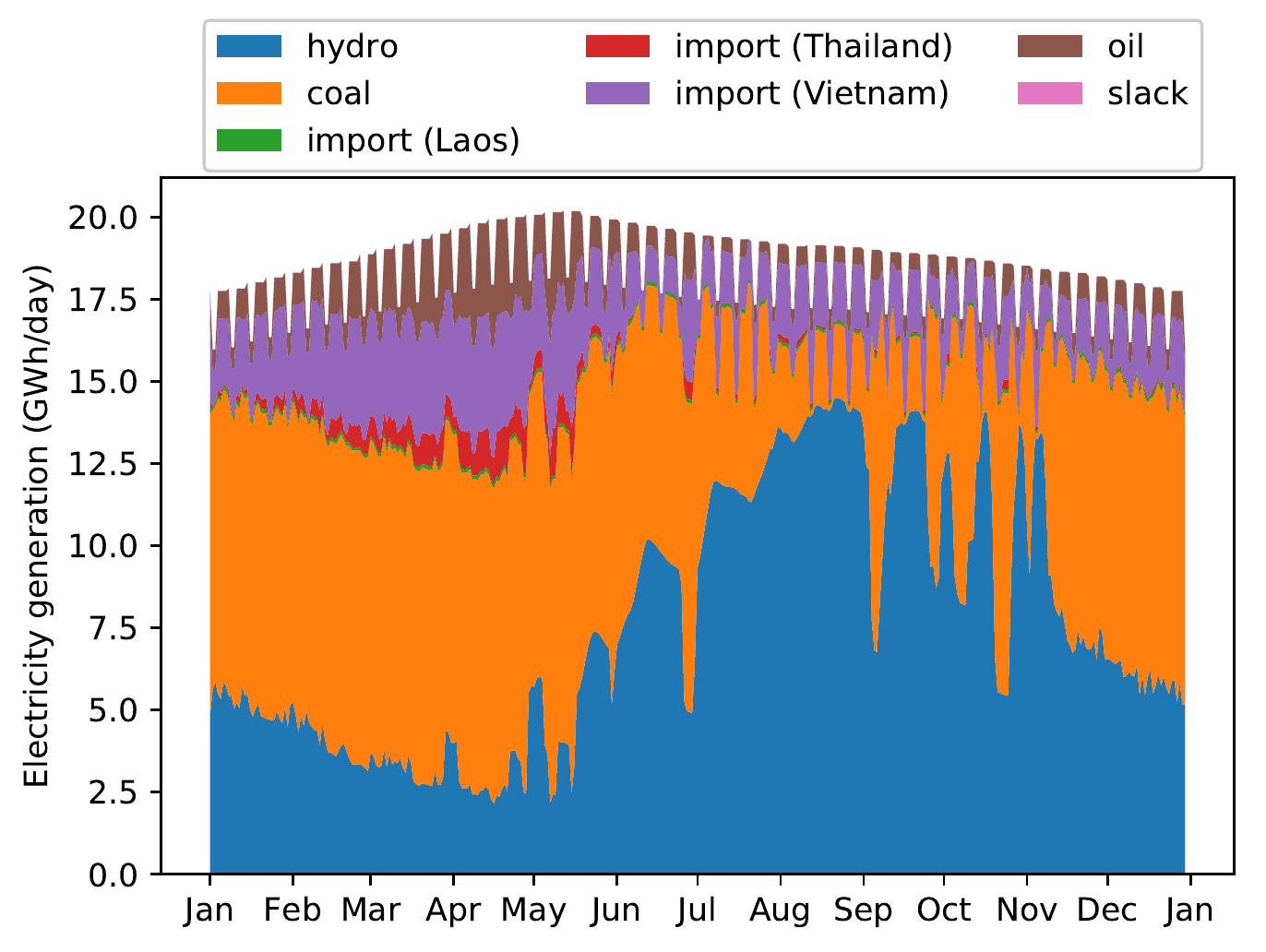}
	\caption{Daily generation mix for 2016.}		
	\label{fig:genmix}
\end{figure}

The simulation for the year of 2016 provides the following hourly time series: (1) operational status of dispatchable units, (2) generation provided by dispatchable units and variable renewable resources, (3) voltage angles at each node, and (4) spinning and non-spinning reserves. These variables represent an important information for the operation, planning, and management of power systems. For example, detailed information on the operational status can support the unit commitment and economic dispatch over the planning horizon, while statistics on the generation mix provide an overview of the system's dependance on the various energy sources. These concepts are further exemplified in Figure \ref{fig:genmix}, which illustrates the energy generation mix. Here, we can notice the `signature' of wet and dry seasons: during the monsoon (May to November), the system heavily relies on hydropower production, while during the pre- and post-monsoon months the electricity generation is largely based on coal and import. The use of these two resources is not only controlled by economic factors---coal is cheaper than imported electricity---but also by the several techno-economic constraints described in Section \ref{constraints}. For example, there are a few days in August in which the concomitance of limited load and high fixed and start-up costs make imported electricity preferable over coal. The above analyses could be readily extended to longer simulation periods or multiple demand patterns, so as to explore the effect of different drivers on the performance of the power system (e.g., \cite{SU2017172,Voisin2016}).\\

\begin{figure}[]
	\centering
	\includegraphics[width=0.75\linewidth]{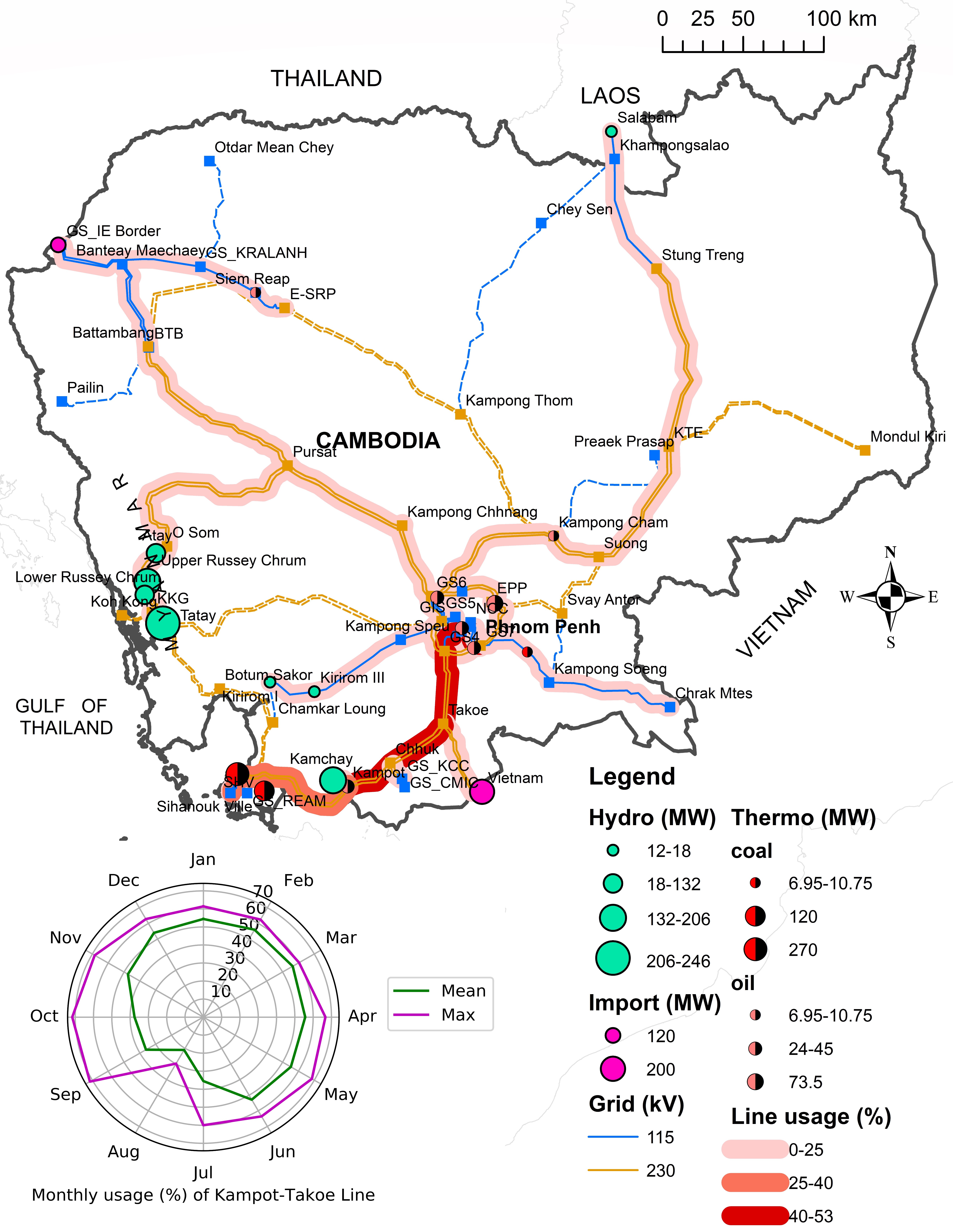}
	\caption{Annual average usage of the transmission lines, expressed as a percentage of the lines' capacity. The radar chart shows the monthly average and maximum usage of the line connecting Kampot to Takoe (one of the lines with relatively high usage).}
	\label{fig:cambgrid_usage}
\end{figure}

Apart from the detailed representation of the aforementioned variables, another feature of PowNet is the explicit representation of the hourly transmission of electricity through the lines (estimated from the voltage angle at each node). This information is synthesized in Figure \ref{fig:cambgrid_usage}, which depicts the annual average usage of each line. Results indicate that usage is limited, with only a few lines showing values larger than 40\% of their capacity. Unsurprisingly, these lines connect the coal plants in the Southwestern coast and import station from Vietnam to the capital (Phnom Penh), namely the area with the highest load. The radar chart indicates that the average usage of these lines is higher during the dry months, when the system heavily relies on the coal units due to the low hydropower availability. Despite the limited average usage, the maximum usage of the lines during wet months is high ($\sim$ 70\%), mainly because of the peak-hour supply from the import station. Note that the transmission of electricity through the lines can also be used to calculate the number of N-1 violations (not observed in this example), a proxy of the network's stress conditions. \\ 

\begin{figure}[h]
	\centering
	\includegraphics[trim={0 0.15cm 0 0.1cm},clip=true,width=0.8\linewidth]{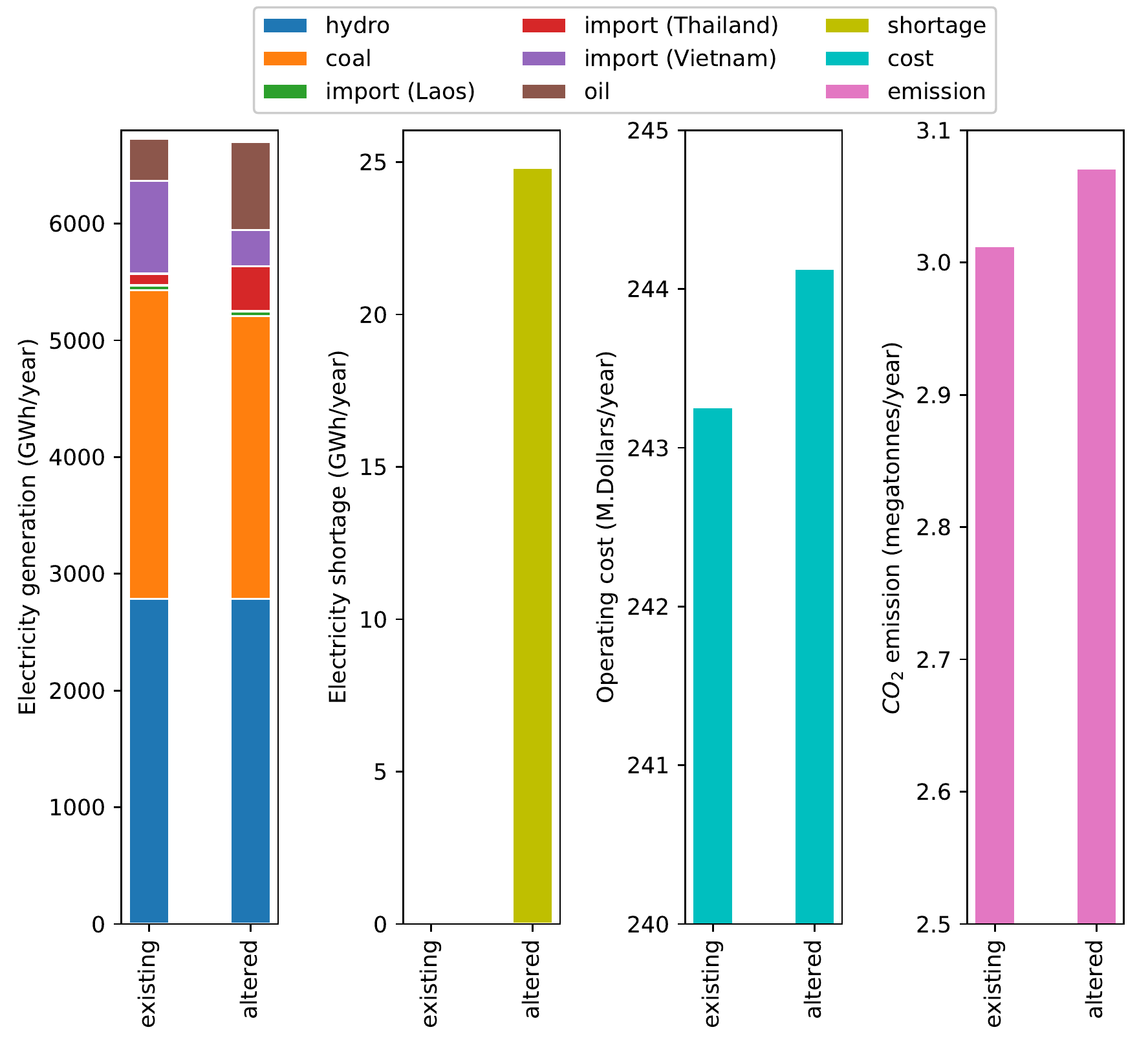}
	\caption{Simulation results obtained with the `existing' and `altered' transmission networks. System's performance is measured in terms of generation mix, electricity shortage, operating costs, and CO$_{2}$ emissions.}
	\label{fig:existing_vs_altered}
\end{figure}

To further demonstrate the importance of a detailed representation of the high-voltage transmission lines, we run PowNet using a slightly altered network. In the `altered' network, we assume that the capacity of the two lines transferring electricity from the Southwestern coast and Vietnam to the capital (see Figure \ref{fig:cambgrid_usage}) is halved---these are double circuit transmission lines, so such scenario represents an instance in which one of the two circuits fails to operate. (All other parameters and input variables remain unchanged.) Figure \ref{fig:existing_vs_altered} compares the system's performance obtained with the `existing' and `altered' transmission networks. Looking at the generation mix (left panel), we note that the model with altered network dispatches less electricity from coal plants and Vietnam's import station. That is because the two lines with reduced capacity are frequently stressed (about 18\% of the time), so they do not allow the coal plants and import station to run at their full capacity. The reduced dispatch from the coal plants and Vietnam's import station is partially offset by higher production from the oil-fired units and Thai import station (we also observe a shortage of electricity supply of about 25 GWh). In turn, such higher reliance on oil affects operating costs and CO$_{2}$ emissions. Overall, these results indicate that the altered functioning of just a few lines can influence the behaviour of an entire power system. Capturing such dynamics is therefore paramount to supporting effectively the operations and planning of large-scale energy systems.

\section{Conclusions}

PowNet is a least-cost optimization model for simulating the UC/ED of national and regional power systems. The model accounts for the techno-economic constraints of generating units and DC transmission network, thereby providing functionalities for both grid and economic analyses. In particular, PowNet is conceived for applications in the water-energy nexus domain, and it can easily incorporate information on the effect of water availability on renewable and non-renewable resources. The data requirements for the DC transmission network are limited, thereby making PowNet applicable to data-scarce regions, particularly developing countries.

\section*{(2) Availability}
\vspace{0.5cm}
\subsection*{\normalsize Operating system}
Windows 10, Linux, and any other operating systems running Python and a standard solver (e.g., Gurobi, CPLEX).

\subsection*{\normalsize Programming language}
Python 3.5

\subsection*{\normalsize Additional system requirements}
None.

\subsection*{\normalsize Dependencies}
PowNet is written in Python 3.5. The following Python packages are required for PowNet: (i) Pyomo, (ii) NumPy, (iii) Pandas, and (iv) Matplotlib (optional for plotting). The Jupyter Notebook (freely available as part of Anaconda Python with the aforementioned packages) is required to run the script `PowNetSolver.ipynb', which also requires an optimization solver (e.g., Gurobi, CPLEX).


\subsection*{\normalsize Software location:}


\begin{description}[noitemsep,topsep=0pt]
	\item[\small Name:] Zenodo 
	\item[\small Persistent identifier:] https://doi.org/10.5281/zenodo.3462879 
	\item[\small Licence:] MIT 
	\item[\small Publisher:] AFM Kamal Chowdhury 
	\item[\small Version published:] v1.1 
	\item[\small Date published:] 27-September-2019 
\end{description}

{\bf Code repository} 

\begin{description}[noitemsep,topsep=0pt]
	\item[\small Name:] GitHub 
	\item[\small Persistent identifier:] https://github.com/kamal0013/PowNet 
	\item[\small Licence:] MIT 
	\item[\small Date published:] 27-September-2019 
\end{description}

%

\subsection*{\normalsize Language}
English

\section*{(3) Reuse potential}


PowNet is available on GitHub with step-by-step instructions on how to formulate and implement the model of a given power system. While the data and scripts demonstrate its use for the Cambodian power system, the user can easily build a new model or customize an existing one by modifying (or adding) dispatchable generators, renewable resources, substations, and transmission lines. The GitHub repository also includes a few additional scripts that help users perform some standard analyses on PowNet's output. Note that Pyomo is free, and that academic users can obtain a free license of state-of-the-art standard solvers (e.g., Gurobi, CPLEX) to solve the model. We also note that PowNet has been tested on Windows and Linux operating systems. Overall, the availability of a dedicated repository, instructions, and extensive examples should make PowNet easy and straightforward to reuse.

\section*{\normalsize Acknowledgment/Funding statement}

This research is supported by Singapore's Ministry of Education (MoE) through the Tier 2 project `Linking water availability to hydropower supply---an engineering systems approach' (Award No. MOE2017-T2-1-143).

\section*{\normalsize Competing interests}

The authors declare that they have no competing interests.


\section*{Appendix}

\section*{\normalsize A. Estimation of transmission parameters}
PowNet uses two parameters for each transmission line, namely susceptance (\textit{LineSus}) and capacity (\textit{LineCap}). For each transmission line, \textit{LineCap} is estimated as the product between the capacity per circuit and number of circuits in the line. \textit{LineSus} is estimated as the reciprocal of line reactance \textit{X}, which is calculated as follows \cite{Bergen1999}:  
\begin{align}\label{eqreac}\tag{A1}
\begin{split}
& {X} = 4 \pi f L \times {10}^{-7} \times \ln \Big( \dfrac{{D}_{m}}{GMR} \Big),
\end{split}
\end{align}	
where, \textit{f} and $L$ are the line frequency (Hz) and length (km), \textit{D\textsubscript{m}} the geometric mean distance between phases, and \textit{GMR} the geometric mean radius of line per phase. Assuming that the lines are in a three-phase system with symmetrically arranged conductors, the geometric mean distance $D_{m}$ can be considered equal to the centre-to-centre distance $D$ (m) between any two conductors \cite{Conejo2018}. $D$ is calculated as a function of the voltage level (kV), as recommended by \cite{Barthold1978,MacCarthy1968}:
\begin{align}\label{eqGeomDist}\tag{A2}
\begin{split}
& {D}_{m} = D = 0.0348 \times \big(0.077 \times Voltage - 3.11\big).
\end{split}
\end{align}	
	
The GMR associated with line is calculated as follows \cite{Conejo2018}:
\begin{align}\label{eqGeomDist}\tag{A3}
\begin{split}
& GMR = \Big( \prod_{c=1}^{C} \big({e}^{-1/4} \times r_{j} \times \prod_{j=1,j\neq c}^{C} {d}_{c,j} \big)^{1/C} \Big)^{1/C},
\end{split}
\end{align}	
where, $r_{j}$ is the radius of any conductor \textit{j}, and $d_{c,j}$ is the centre-to-centre distance between \textit{j} and each of the other conductors of a phase of \textit{C} conductors.\\

For the Cambodian power system, we obtained data on voltage level (kV), number of circuits, line length (km), size (mm\textsuperscript{2}) and number of conductors, and capacity (MW) per circuit from \cite{EDC2016,JICA2014}.

\section*{\normalsize B. Simulation of hydropower availability}

Table B1 shows the design specifications (collected from \cite{IR2008,MEN201440}) of the Cambodian hydropower reservoirs. To calculate the hydropower available in each dam, we adopted the following procedure. First, we estimated the inflow to the reservoirs (${Q}_{t}$) through the use of the rational method, that is:
\begin{align}\label{hydro1}\tag{B1}
\begin{split}
& {Q}_{t} = c \times {I}_{t} \times A,
\end{split}
\end{align}	
where $c$ is a runoff coefficient, ${I}_{t}$ the rainfall intensity (mm/day), and $A$ the drainage area of each dam (km$^2$). The runoff coefficient was estimated as a function of the soil type and slope of the drainage basins, while the rainfall intensity was extracted from the APHRODITE gridded rainfall dataset. \\

Then, we calculated the mass balance of each reservoir, namely:
\begin{align}\label{hydro2}\tag{B2}
\begin{split}
& {S}_{t+1} = {S}_{t} + {Q}_{t} - {R}_{t},
\end{split}
\end{align}	
\begin{align}\label{hydro3}\tag{B4}
\begin{split}
\text{s.t. } & 0 \leq S_t \leq {S}_{cap},
\end{split}
\end{align}	
where ${S}_{t}$, $Q_t$, and $R_t$ represent the water storage, inflow, and release at time $t$. $S_{cap}$ is the capacity of the reservoir. To calculate the release $R_t$, we adopted rule curves aimed to maximize the hydropower production \cite{Piman2012}. \\ 

Finally, we used the hydropower equation to estimate the power available at each hydropower plant:
\begin{align}\label{hydro5}\tag{B5}
\begin{split}
& {hydropower}_{t} = \eta \times \rho \times g \times {R}_{t} \times {H}_{t},
\end{split}
\end{align}	
where ${hydropower}_{t}$ is the available hydropower (MW), $\eta$ a non-dimensional turbine efficiency term, $\rho$ the water density (1,000 kg/m\textsuperscript{3}), $g$ the gravitational acceleration (9.81 m/s\textsuperscript{2}), and ${H}_{t}$ the hydraulic head at time $t$ (calculated from the storage $S_t$). The daily time series of available hydropower was transformed into an hourly time series by assuming equal water availability throughout a single day.

\begin{table}[h]
	\centering
	\renewcommand\thetable{B1}
	\caption{Design specifications of the Cambodian hydropower dams.}
	\label{TabDam}
	\begin{tabular}{lccccccc}
		\hline
		Name        & \begin{tabular}[c]{@{}c@{}}Installed \\ capacity\end{tabular} & \begin{tabular}[c]{@{}c@{}}Dam \\ height\end{tabular} & Storage   & \begin{tabular}[c]{@{}c@{}}Design \\ discharge\end{tabular} & \begin{tabular}[c]{@{}c@{}}Hydraulic \\ head\end{tabular} & \begin{tabular}[c]{@{}c@{}}Basin \\ area\end{tabular} & \begin{tabular}[c]{@{}c@{}}Runoff \\ coefficient\end{tabular} \\
		& (MW)                                                        & (m)                                                   & (Mm$^3$) & (m$^3$/s)                                                      & (m)                                                   & (km$^2$)                                                 &                                                               \\ \hline
		Kamchay     & 194.1                                                       & 110                                                   & 680       & 163.5                                                       & 122                                                   & 710                                                   & 0.62                                                          \\
		Kirirom I   & 12                                                          & 34                                                    & 30        & 20                                                          & 373.5                                                 & 99                                                    & 0.48                                                          \\
		Kirirom III & 18                                                          & 40                                                    & 30        & 40                                                          & 271                                                   & 105                                                   & 0.48                                                          \\
		L.R. Chrum  & 338                                                         & 68                                                    & 62        & 300                                                         & 132                                                   & 1,550                                                 & 0.51                                                          \\
		Stung Atay  & 240                                                         & 45                                                    & 443.8     & 125                                                         & 216                                                   & 1,157                                                 & 0.75                                                          \\
		Stung Tatay & 246                                                         & 77                                                    & 322       & 150                                                         & 188                                                   & 1,073                                                 & 0.71                                                          \\ \hline
	\end{tabular}
\end{table}

\section*{References}
\renewcommand{\section}[2]{}%
\bibliography{pownet}

\begin{thebibliography}{38}
\providecommand{\natexlab}[1]{#1}
\providecommand{\url}[1]{\texttt{#1}}
\expandafter\ifx\csname urlstyle\endcsname\relax
  \providecommand{\doi}[1]{doi: #1}\else
  \providecommand{\doi}{doi: \begingroup \urlstyle{rm}\Url}\fi

\bibitem[Gielen et~al.(2019)Gielen, Boshell, Saygin, Bazilian, Wagner, and
  Gorini]{GIELEN201938}
Dolf Gielen, Francisco Boshell, Deger Saygin, Morgan~D. Bazilian, Nicholas
  Wagner, and Ricardo Gorini.
\newblock The role of renewable energy in the global energy transformation.
\newblock \emph{Energy Strategy Reviews}, 24:\penalty0 38 -- 50, 2019.

\bibitem[Welsch et~al.(2014)Welsch, Deane, Howells, Gallach{\'o}ir, Rogan,
  Bazilian, and Rogner]{WELSCH2014600}
Manuel Welsch, Paul Deane, Mark Howells, Brian~{\'O} Gallach{\'o}ir, Fionn
  Rogan, Morgan Bazilian, and Hans-Holger Rogner.
\newblock Incorporating flexibility requirements into long-term energy system
  models--{A} case study on high levels of renewable electricity penetration in
  {I}reland.
\newblock \emph{Applied Energy}, 135:\penalty0 600--615, 2014.

\bibitem[Su et~al.(2017)Su, Kern, and Characklis]{SU2017172}
Yufei Su, Jordan~D. Kern, and Gregory~W. Characklis.
\newblock The impact of wind power growth and hydrological uncertainty on
  financial losses from oversupply events in hydropower-dominated systems.
\newblock \emph{Applied Energy}, 194:\penalty0 172--183, 2017.

\bibitem[Ringkjøb et~al.(2018)Ringkjøb, Haugan, and
  Solbrekke]{RINGKJOB2018440}
Hans-Kristian Ringkjøb, Peter~M. Haugan, and Ida~Marie Solbrekke.
\newblock A review of modelling tools for energy and electricity systems with
  large shares of variable renewables.
\newblock \emph{Renewable and Sustainable Energy Reviews}, 96:\penalty0 440 --
  459, 2018.

\bibitem[Brown et~al.(2018)Brown, H{\"o}rsch, and
  Schlachtberger]{brown2018pypsa}
Thomas Brown, Jonas H{\"o}rsch, and David Schlachtberger.
\newblock {P}y{P}{S}{A}: {P}ython for {P}ower {S}ystem {A}nalysis.
\newblock \emph{Journal of Open Research Software}, 6\penalty0 (1), 2018.

\bibitem[Dai et~al.(2018)Dai, Wu, Han, Weinberg, Xie, Wu, Song, Jia, Xue, and
  Yang]{DAI2018393}
Jiangyu Dai, Shiqiang Wu, Guoyi Han, Josh Weinberg, Xinghua Xie, Xiufeng Wu,
  Xingqiang Song, Benyou Jia, Wanyun Xue, and Qianqian Yang.
\newblock Water-energy nexus: A review of methods and tools for
  macro-assessment.
\newblock \emph{Applied Energy}, 210:\penalty0 393--408, 2018.

\bibitem[van Vliet et~al.(2016)van Vliet, Wiberg, Leduc, and
  Riahi]{vanVliet_2016}
Michelle T.~H. van Vliet, David Wiberg, Sylvain Leduc, and Keywan Riahi.
\newblock Power-generation system vulnerability and adaptation to changes in
  climate and water resources.
\newblock \emph{Nature Climate Change}, 6:\penalty0 375--380, 2016.

\bibitem[Turner et~al.(2017)Turner, Ng, and Galelli]{TURNER2017663}
Sean~W.D. Turner, Jia~Yi Ng, and Stefano Galelli.
\newblock Examining global electricity supply vulnerability to climate change
  using a high-fidelity hydropower dam model.
\newblock \emph{Science of The Total Environment}, 590-591:\penalty0 663--675,
  2017.

\bibitem[Turner et~al.(2019)Turner, Voisin, Fazio, Hua, and
  Jourabchi]{Turner2019}
S.~W.~D. Turner, N.~Voisin, J.~Fazio, D.~Hua, and M.~Jourabchi.
\newblock Compound climate events transform electrical power shortfall risk in
  the {P}acific {N}orthwest.
\newblock \emph{Nature Communications}, 10:\penalty0 1--8, 2019.

\bibitem[Conejo and Baringo(2018)]{Conejo2018}
A.J. Conejo and L.~Baringo.
\newblock \emph{Power system operations}.
\newblock Springer, 1st. edition, 2018.

\bibitem[Voisin et~al.(2016)Voisin, Kintner-Meyer, Skaggs, Nguyen, Wu, Dirks,
  Xie, and Hejazi]{Voisin2016}
N.~Voisin, M.~Kintner-Meyer, R.~Skaggs, T.~Nguyen, D.~Wu, J.~Dirks, Y.~Xie, and
  M.~Hejazi.
\newblock Vulnerability of the {U.S.} western electric grid to
  hydro-climatological conditions: How bad can it get?
\newblock \emph{Energy}, 115:\penalty0 1--12, 2016.

\bibitem[Deane et~al.(2012)Deane, Chiodi, Gargiulo, and
  Gallachóir]{DEANE2012303}
J.P. Deane, Alessandro Chiodi, Maurizio Gargiulo, and Brian P.~Ó Gallachóir.
\newblock Soft-linking of a power systems model to an energy systems model.
\newblock \emph{Energy}, 42\penalty0 (1):\penalty0 303--312, 2012.

\bibitem[Savvidis et~al.(2019)Savvidis, Siala, Weissbart, Schmidt, Borggrefe,
  Kumar, Pittel, Madlener, and Hufendiek]{SAVVIDIS2019503}
Georgios Savvidis, Kais Siala, Christoph Weissbart, Lukas Schmidt, Frieder
  Borggrefe, Subhash Kumar, Karen Pittel, Reinhard Madlener, and Kai Hufendiek.
\newblock The gap between energy policy challenges and model capabilities.
\newblock \emph{Energy Policy}, 125:\penalty0 503 -- 520, 2019.

\bibitem[{Milano}(2005)]{PSAT}
F.~{Milano}.
\newblock An open source power system analysis toolbox.
\newblock \emph{IEEE Transactions on Power Systems}, 20\penalty0 (3):\penalty0
  1199--1206, 2005.

\bibitem[O'Connell et~al.(2019)O'Connell, Voisin, Macknick, and
  Fu]{OCONNELL2019745}
O'Connell, Nathalie Voisin, Macknick, and Fu.
\newblock Sensitivity of western u.s. power system dynamics to droughts
  compounded with fuel price variability.
\newblock \emph{Applied Energy}, 247:\penalty0 745 -- 754, 2019.

\bibitem[Kern and Characklis(2017)]{kern2017}
Jordan~D. Kern and Gregory~W. Characklis.
\newblock Evaluating the financial vulnerability of a major electric utility in
  the {S}outheastern {U.S.} to drought under climate change and an evolving
  generation mix.
\newblock \emph{Environmental Science \& Technology}, 51\penalty0
  (15):\penalty0 8815--8823, 2017.

\bibitem[{Stott} et~al.(2009){Stott}, {Jardim}, and {Alsac}]{Stott2009}
B.~{Stott}, J.~{Jardim}, and O.~{Alsac}.
\newblock {D}{C} power flow revisited.
\newblock \emph{IEEE Transactions on Power Systems}, 24\penalty0 (3):\penalty0
  1290--1300, 2009.

\bibitem[{Brown} et~al.(2016){Brown}, {Schierhorn}, {Tröster}, and
  {Ackermann}]{Brown2016}
T.~{Brown}, P.~{Schierhorn}, E.~{Tröster}, and T.~{Ackermann}.
\newblock Optimising the european transmission system for 77\% renewable
  electricity by 2030.
\newblock \emph{IET Renewable Power Generation}, 10\penalty0 (1):\penalty0
  3--9, 2016.

\bibitem[Schlecht and Weigt(2014)]{Schlecht20141}
Ingmar Schlecht and Hannes Weigt.
\newblock Swissmod - a model of the {S}wiss electricity market.
\newblock \emph{FoNEW Discussion Paper}, page~31, 2014.

\bibitem[Truby(2014)]{Truby2014}
Johannes Truby.
\newblock Thermal power plant economics and variable renewable energies: A
  model-based case study for {G}ermany.
\newblock Technical report, International Energy Agency (IEA), 2014.

\bibitem[Lubega and Stillwell(2018)]{lubega2018maintaining}
William~Naggaga Lubega and Ashlynn~S Stillwell.
\newblock Maintaining electric grid reliability under hydrologic drought and
  heat wave conditions.
\newblock \emph{Applied Energy}, 210:\penalty0 538--549, 2018.

\bibitem[Dang et~al.(2019)Dang, Chowdhury, and Galelli]{Dang2019}
T.~D. Dang, A.~K. Chowdhury, and S.~Galelli.
\newblock On the representation of water reservoir storage and operations in
  large-scale hydrological models: implications on model parameterization and
  climate change impact assessments.
\newblock \emph{Hydrology and Earth System Sciences Discussions},
  2019:\penalty0 1--34, 2019.
\newblock \doi{10.5194/hess-2019-334}.

\bibitem[{Papavasiliou} et~al.(2015){Papavasiliou}, {Oren}, and
  {Aravena}]{Papavasiliou}
A.~{Papavasiliou}, S.~S. {Oren}, and I.~{Aravena}.
\newblock Stochastic modeling of multi-area wind power production.
\newblock In \emph{2015 48th Hawaii International Conference on System
  Sciences}, pages 2616--2626, 2015.

\bibitem[Blair et~al.(2014)Blair, Dobos, Freeman, Neises, Wagner, Ferguson,
  Gilman, and Janzou]{SAM2014}
N.~Blair, A.~Dobos, J.~Freeman, T.~Neises, M.~Wagner, T.~Ferguson, P.~Gilman,
  and S.~Janzou.
\newblock {S}ystem advisor model, {SAM} 2014.1.14: General description, 2014.
\newblock url: https://sam.nrel.gov, last accessed on 05/09/2019.

\bibitem[Bergen and Vittal(1999)]{Bergen1999}
Arthur~R. Bergen and Vijay. Vittal.
\newblock \emph{Power systems analysis}.
\newblock Prentice Hall, Upper Saddle River, New Jersey, 2nd. edition, 1999.

\bibitem[Guerra et~al.(2016)Guerra, Tejada, and Reklaitis]{GUERRA20161}
Omar~J. Guerra, Diego~A. Tejada, and Gintaras~V. Reklaitis.
\newblock An optimization framework for the integrated planning of generation
  and transmission expansion in interconnected power systems.
\newblock \emph{Applied Energy}, 170:\penalty0 1--21, 2016.

\bibitem[Hart et~al.(2011)Hart, Paul, and David]{Hart2011}
W~E Hart, W~J Paul, and L~W David.
\newblock Pyomo: Modeling and solving mathematical programs in {P}ython.
\newblock \emph{Mathematical Programming Computation}, 3:\penalty0 219--260,
  2011.

\bibitem[Chowdhury et~al.(2018)Chowdhury, Dang, and Galelli]{PowNetLaos1}
A.~K. Chowdhury, T.~D. Dang, and S.~Galelli.
\newblock {C}oupling hydrologic and network constrained unit commitment models
  to understand the water-energy nexus in {L}aos, 2018.
\newblock {P}resented in AGU Fall Meeting 2018, Washington D.C. url:
  http://adsabs.harvard.edu/abs/2018AGUFM.H23F..03C.

\bibitem[Galelli et~al.(2019)Galelli, Chowdhury, and Dang]{PowNetLaos2}
S.~Galelli, A.~K. Chowdhury, and T.~D. Dang.
\newblock {A} coupled water-energy model reveals key interdependencies between
  hydro-climatic variability, energy generation, and power distribution in the
  {G}reater {M}ekong {S}ub-region, 2019.
\newblock {P}resented in EGU General Assembly 2019, Vienna. url:
  https://meetingorganizer.copernicus.org/EGU2019/EGU2019-6293.pdf.

\bibitem[Chowdhury et~al.(2019)Chowdhury, Dang, and Galelli]{PowNetThai}
A.~K. Chowdhury, T.~D. Dang, and S.~Galelli.
\newblock {I}mpacts of hydro-climatic variability on the energy system of the
  {G}reater {M}ekong {S}ub-region, 2019.
\newblock {P}resented in AOGS Annual Meeting 2019, Singapore.

\bibitem[EDC(2016)]{EDC2016}
EDC.
\newblock Annual report 2016.
\newblock Technical report, Electricite Du Cambodge (EDC), 2016.

\bibitem[EIA(2016)]{EIA2016}
EIA.
\newblock Average tested heat rates by prime mover and energy source, 2007 -
  2016 ({F}orm {EIA}-860, annual electric generator report).
\newblock Technical report, U.S. Energy Information Administration (EIA), 2016.

\bibitem[Barthold et~al.(1978)Barthold, Clayton, Grant, Longo, Stewart, and
  Wilson]{Barthold1978}
L.O. Barthold, R.E. Clayton, I.S. Grant, V.J. Longo, J.R. Stewart, and D.D.
  Wilson.
\newblock Transmission line reference book: 115–138 k{V} compact line design.
\newblock Technical report, Electric Power Research Institute, Inc., Palo Alto,
  CA, 1978.

\bibitem[MacCarthy(1968)]{MacCarthy1968}
D.D. MacCarthy.
\newblock \emph{{EHV} Transmission Line Reference Book}.
\newblock Project EHV, General Electric Company, Edison Electric Institute,
  1968.

\bibitem[JICA(2014)]{JICA2014}
JICA.
\newblock Final report: Preparatory survey for {P}hnom {P}enh city transmission
  and distribution system expansion project phase ii in the {K}ingdom of
  {C}ambodia.
\newblock Technical report, Japan International Cooperation Agency (JICA),
  2014.

\bibitem[Middleton and Chanthy(2008)]{IR2008}
C.~Middleton and S.~Chanthy.
\newblock {C}ambodia’s hydropower development and {C}hina’s involvement.
\newblock Technical report, {I}nternational {R}ivers, 2008.

\bibitem[Men et~al.(2014)Men, Thun, Yin, and Lebel]{MEN201440}
Prachvuthy Men, Vathana Thun, Soriya Yin, and Louis Lebel.
\newblock Benefit sharing from {K}amchay and {L}ower {S}esan 2 hydropower
  watersheds in {C}ambodia.
\newblock \emph{Water Resources and Rural Development}, 4:\penalty0 40 -- 53,
  2014.

\bibitem[Piman et~al.(2013)Piman, Cochrane, Arias, Green, and Dat]{Piman2012}
T.~Piman, T.~A. Cochrane, M.~E. Arias, A.~Green, and N.~D. Dat.
\newblock Assessment of flow changes from hydropower development and operations
  in {S}ekong, {S}esan, and {S}repok {R}ivers of the {M}ekong basin.
\newblock \emph{Journal of Water Resources Planning and Management},
  139\penalty0 (6):\penalty0 723--732, 2013.

\end{thebibliography}





\end{document}